\documentclass[iop]{emulateapj}

\usepackage{graphics}
\usepackage{mathptmx}
\usepackage{natbib}

\newcommand{\teff}{$T_{\!\mbox{\scriptsize\em eff}}$}

\newcommand{\msun}{$M_\odot$}

\def\kpc{\mbox{${\rm kpc}^{-1}$}}
 \newcommand{\hii}{\ion{H}{2}}

\slugcomment{To appear in Astrophysical Journal}

\shorttitle{Stellar Metallicity of NGC4258}
\shortauthors{Kudritzki et al.}

\begin{document}


\title{A Direct Stellar Metallicity Determination in the Disk of the Maser Galaxy NGC4258}


\author{Rolf-Peter Kudritzki\altaffilmark{1,2,3},
Miguel A. Urbaneja\altaffilmark{4},
J. Zachary Gazak\altaffilmark{1},
Lucas Macri\altaffilmark{5},
Matthew W. Hosek Jr.\altaffilmark{1},
Fabio Bresolin\altaffilmark{1},
Norbert Przybilla\altaffilmark{4}
}

\begin{abstract}

  We present the first direct determination of a stellar metallicity in the
  spiral galaxy NGC\,4258 ($D=7.6$\,Mpc) based on the quantitative analysis of
  a low-resolution ($\sim 5$ \AA) Keck LRIS spectrum of a blue supergiant star
  located in its disk. A determination of stellar metallicity in this galaxy is
  important for the absolute calibration of the Cepheid Period-Luminosity
  relation as an anchor for the extragalactic distance scale and for a better
  characterization of its dependence as a function of abundance.  We find a
  value 0.2 dex lower than solar metallicity at a galactocentric distance of
  8.7\,kpc, in agreement with recent \hii~region studies using the weak forbidden
  auroral oxygen line at 4363~\AA. We determine the effective stellar
  temperature, gravity, luminosity and line-of-sight extinction of the blue 
  supergiant being studied. We show that it fits well on the flux-weighted
  gravity--luminosity relation (FGLR), strengthening the potential of this method
  as a new extragalactic distance indicator.

\end{abstract}



\altaffiltext{1}{Institute for Astronomy, University of Hawai'i, 2680 Woodlawn Dr, Honolulu, HI 96822, USA}
\altaffiltext{2}{University Observatory Munich, Scheinerstr. 1, D-81679 Munich, Germany}
\altaffiltext{3}{Max-Planck-Institute for Astrophysics, Karl-Schwarzschild-Str.1, D-85741 Garching, Germany}
\altaffiltext{4}{Institute for Astro- and Particle Physics, University of Innsbruck, Technikerstr. 25/8, A-6020 Innsbruck, Austria}
\altaffiltext{5}{George P. and Cynthia Woods Mitchell Institute for Fundamental Physics and Astronomy, 
Department of Physics \& Astronomy, Texas A\&M University, 4242 TAMU, College Station, TX77843-4242, USA}



\section{Introduction}

The precise VLBI mapping of water masers orbiting the central black hole in NGC\,4258 makes
it possible to measure a geometrical distance to this spiral galaxy with unprecedented accuracy
\citep{herrnstein99, humphreys08}. Most recently, \citet{humphreys13} have refined these measurements using a new
model which includes disk warping and confocal elliptical maser orbits with differential precession. They obtained a
distance of 7.60$\pm$0.17$\pm$0.15\,Mpc, where the uncertainties are split into formal fitting errors and a
systematic term respectively.

Given such an accurate and precise distance, it is an obvious (though bold) step to use this
  galaxy as a new anchor point for extragalactic distance indicators, such as the Cepheid Period-Luminosity
  Relation \citep[][PLR]{leavitt12}. After the discovery with HST of a large sample of Cepheids by
\citet{macri06}, this galaxy was used by \citet{riess09} and \citet{riess11} as the first step in an
HST survey for Cepheids in host galaxies of SNe~Ia out to 30 Mpc, which resulted in a determination
  of the Hubble constant with a total uncertainty of 3.3\%.

Simultaneously, there has also been a dramatic improvement in the measurement of the distance to the Large
Magellanic Cloud. In a long-term project over eight years, \citet{pietrzynski13} accumulated photometric and
spectroscopic observations of long-period, well-detached late-type eclipsing binaries to construct
high-precision light and radial velocity curves of eight systems distributed along the line of nodes towards the
LMC. The analysis of this unique observational material provided radius, extinction, luminosity and distance for each
object resulting in a LMC distance with an unprecedented accuracy of 2\%.

Combining the LMC and NGC\,4258 as two anchor points of the extragalactic distance scale has the potential to
further increase the robustness of the determination of the Hubble constant or, at least, to check for
additional systematic effects. In the case of the PLR, HST studies of Cepheid
fields at different galactocentric radii within spiral galaxies usually show that the variables in
inner fields seem to be brighter by a few tenths of a magnitude than those in outer fields. This is usually
interpreted as an effect of the metallicity gradient in spiral galaxies and a metallicity dependence of the PLR
\citep{kennicutt98, freedman01, mccommas09, gerke11,shappee11}. The Cepheids detected in the maser galaxy NGC\,4258
also show such an effect \citep{macri06}. According to these studies, a difference between solar neighborhood
metallicity and LMC metallicity of 0.4 dex will have an effect of 0.12 mag in distance modulus or 6\% in distance,
clearly too large if the ultimate goal is to measure a Hubble constant with a precision of 3\% or better to constrain
the equation of state of dark energy parameter $w$ (see \citealt{macri06} or \citealt{riess11}). While the interpretation of
the physical reason of the brightness difference of inner and outer field Cepheids is heavily disputed (see
\citealt{bresolin11}, \citealt{kud12}, \citealt{majaess11}, \citealt{storm11}) and while this effect might be less
pronounced in the near-IR H-band used by \citet{riess11}, it is very evident that a precise determination of distances
using Cepheids should be accompanied by an accurate determination of stellar metallicities.

\begin{figure}[!]
 \begin{center}
  \includegraphics[width=0.45\textwidth]{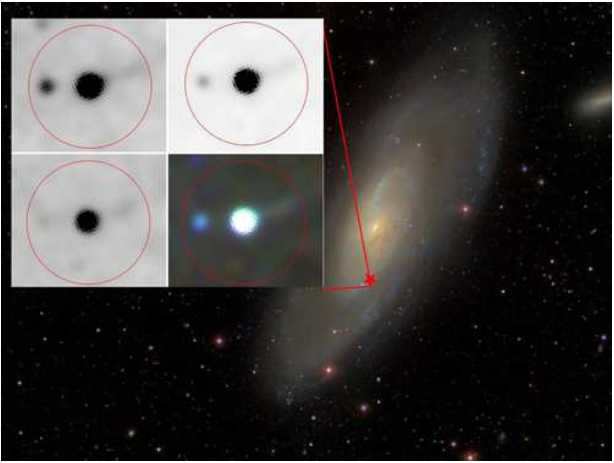}
  \caption[]{
The location of the BSG target in NGC\,4258 together with an enlarged B, V, I and RGN composite HST ACS image. The circle corresponds to 1 arcsec radius. For discussion of the second source see text. \label{target_b} }
 \end{center}
\end{figure}

At distances significantly beyond the Magellanic Clouds, a direct determination of the metallicity of Cepheids based
on spectroscopy is not possible. Instead, the work on the extragalactic distance scale cited above has used the oxygen
abundance obtained from the strong emission lines of \hii~regions as a proxy for stellar metallicities. However, it
has been shown over the last years (see \citealt{kud08}, \citealt{kewley08}, \citealt{bresolin09}, \citealt{u09},
\citealt{kud12} for a detailed discussion) that these ``strong-line methods'' are subject to systematic uncertainties
as large as 0.6 dex, which are poorly understood and can severely affect the inferred values of galaxy
central metallicities and abundance gradients. This introduces an element of uncertainty in the use of Cepheids
as extragalactic distance indicators, in particular, with the goal to obtain very accurate distances.

The important case of the maser galaxy is an illustrative example for this uncertainty. \citet{macri06} used oxygen
abundances obtained from \hii~region strong line studies by \citet{zaritsky94} for their discussion of Cepheid
metallicity in NGC\,4258. These studies claim a very high central value of [O/H]=12+log(O/H)=9.17 and a rather steep
gradient of -0.028 \kpc. On the other hand, the study by \citet{bresolin11} included four \hii~regions
with observations of the weak auroral [OIII] 4363 line, and determined a much lower central metallicity value of
[O/H]=8.49 and a significantly shallower gradient of -0.010 \kpc~(all values are taken from \citealt{bresolin11} but
have been re-normalized to the new maser distance of 7.6 Mpc).  At a galactocentric radius of 8.7\,kpc (see
below) this leads to a striking difference in oxygen abundance. Using the Zaritsky et al. result we obtain [O/H]=8.93,
significantly larger than the solar value of [O/H]=8.69 \citep{allende01,asplund09}, whereas adopting
the Bresolin result yields [O/H]=8.40 which is only slightly larger than the LMC value [O/H]=8.36 (also based on \hii~regions
by \citealt{bresolin11}). Obviously, when discussing Cepheid magnitude differences between NGC\,4258 and the LMC, the
question of whether the metallicity difference is 0.5 dex or almost zero is important.

While at the distance of the maser galaxy a direct spectroscopic investigation of Cepheid metallicities is not
possible to settle this issue, there is an attractive alternative: the quantitative spectroscopy of blue
supergiant stars (BSGs). BSGs are massive stars in the range between 15 and 40~\msun, which cross the
Hertzpsrung-Russell diagram in $10^{4-5}$~yr from the main sequence to the red supergiant stage as stars of spectral
type late B or A. Having an age of $\sim$ 10 million years, they belong to the same population as Cepheids
albeit they are more massive and slightly younger. Their magnitudes can reach up to $M_{V} \cong
-10$, rivaling the integrated brightness of globular clusters and dwarf spheroidal galaxies. They are more
than four magnitudes brighter than Cepheids and, thus, perfect candidates for quantitative stellar abundance studies
beyond the Local Group. \citet{kud08} have shown that accurate metallicities based on elements such as iron, chromium,
titanium etc. can be determined from low resolution spectroscopy of individual objects using refined NLTE model
atmosphere diagnostic methods. This technique has now been applied on a variety of galaxy abundance studies (WLM --
\citealt{bresolin06}; \citealt{urbaneja08}; NGC 3109 -- \citealt{evans07}, \citealt{hosek13}; IC 1613 --
\citealt{bresolin07}; M33 -- \citealt{u09}; NGC 300 -- \citealt{kud08}; M81 -- \citealt{kud12}).

The extension of this technique to the maser galaxy is a consequent next step. So far the most distant galaxy 
studied in the this way has been M81 with a distance of 3.5~Mpc. Investigating BSGs in NGC\,4258 
doubles the range in distance, which is a clear challenge pushing the method to a new limit. We have, thus, 
started a pilot project of multi-object spectroscopy of BSGs in this key galaxy for the extragalactic 
distance scale. In this paper, we present a first result, the direct determination of the metallicity of 
one BSG in NGC\,4258 at a galactocentric distance of 8.7\,kpc.

\section{Observations and Data Reduction}

\begin{figure}
\begin{center}
  \includegraphics[scale=0.3,angle=90]{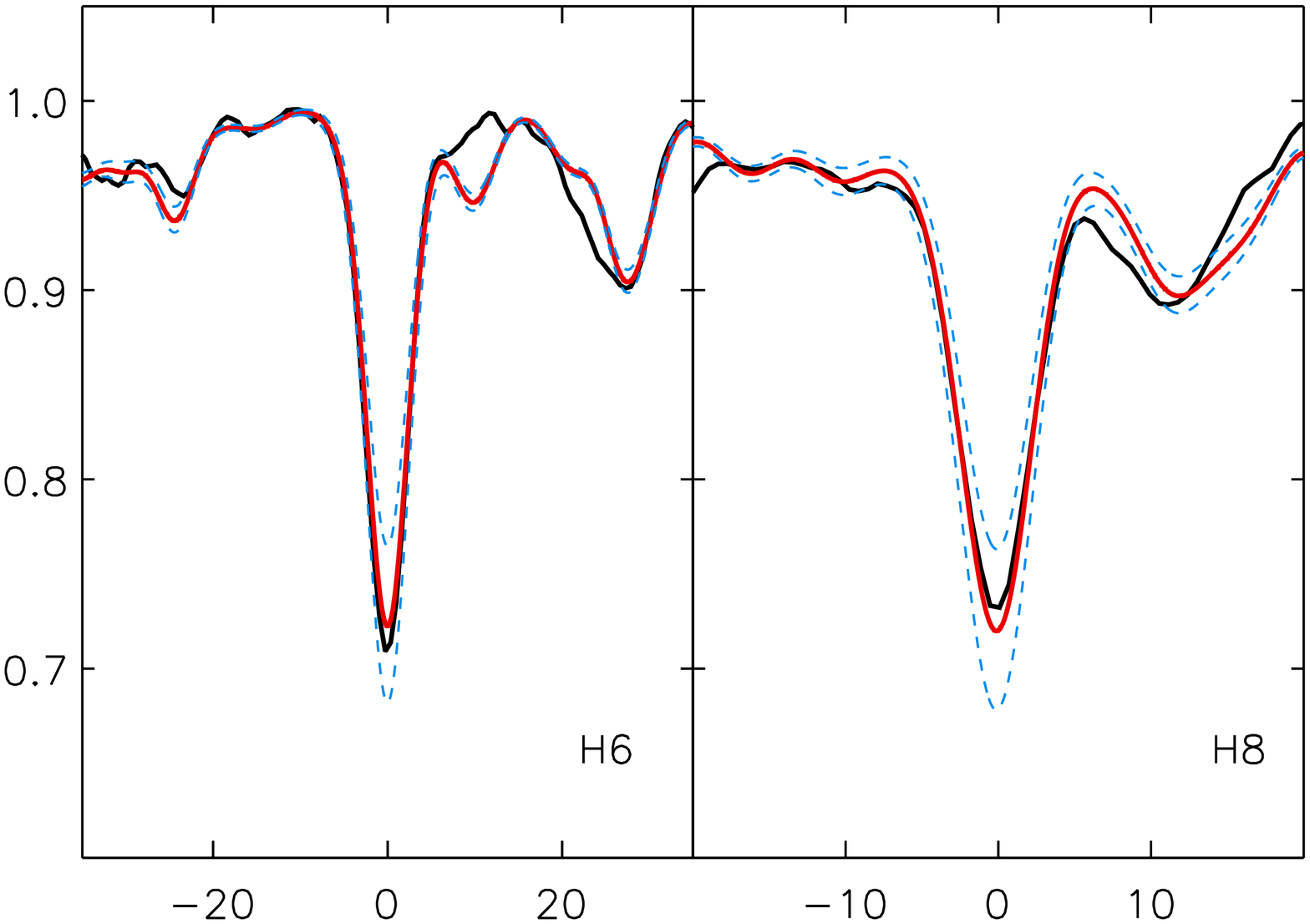}
 \includegraphics[scale=0.3,angle=90]{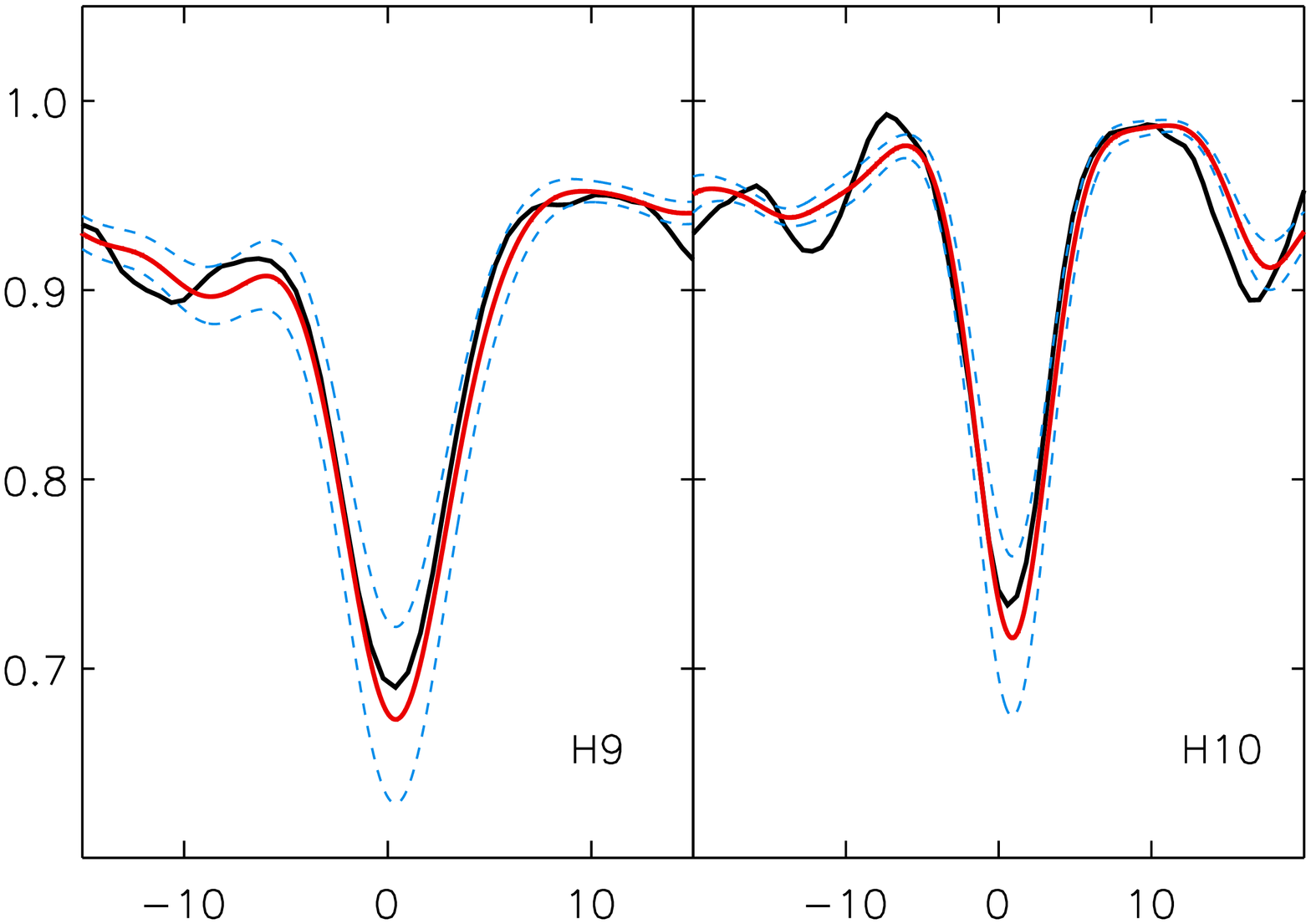} 
\caption{Fit of observed Balmer line profiles (black solid) with model atmospheres of \teff=8300K and 
log g = 0.85 (red, solid) and 0.80 and 0.90 (both blue dashed), respectively. The gravity g is calculated in cgs units.  \label{balmer}}
 \end{center}
\end{figure}

Spectroscopic observations with one multi-object mask were carried out during the night of March 16, 2012, with the
Keck 1 telescope on Mauna Kea and the Low Resolution Imaging Spectrograph \citep[LRIS,][]{oke95} using the atmospheric
dispersion corrector, a slit width of 1.2 arcseconds, the D560 dichroic and the 600/4000 grism (0.63~\AA\,pix$^{-1}$)
and the 900/5500 grating (0.53~\AA\,pix$^{-1}$) in the blue and red channel, respectively. In this paper, we will
discuss and analyze the blue channel (LRIS-B) spectra only, which have a resolution of 5 \AA~FHWM.  One MOS field was
prepared for this night with 23 targets. The BSG candidate targets were selected from an HST/ACS B, V, I survey of
NGC\,4258 (HST-GO-10802 and HST-GO-11570, P.I.: Adam Riess) that covers most of the disk of the galaxy in
17 fields. PSF photometry of all fields yielded 240 BSG candidates with -0.1 mag $\le$ B-V $\le$ 0.5 mag
and V $\le$ 22.0 mag (corresponding to M$_V \le$-7.9 mag). Each candidate target was carefully inspected with regard
to multiplicity. Targets for the mask were selected in a magnitude range from V = 20.4 to 22.0 mag to allow for a
independent determination of distance using the FGLR-method (see below).

Unfortunately, the Keck observations were compromised by poor seeing of 1.3 arcsec (and sometimes worse) throughout
the night. We have, therefore, restricted our investigation to the analysis of the brightest target (slit 6) with
coordinates $\alpha$(2000)=184.75836 and $\delta$(2000)=+47.26312 and the following photometric properties:
V=$20.58\pm0.02$~mag , B-V=$0.29\pm0.05$~mag, V-I=$0.48\pm0.03$~mag in the Johnson-Cousins system. The deprojected
galactocentric distance in units of the R$_{25}$ radius is R/R$_{25}$=0.42 or (with R$_{25}$=20.6\,kpc at a distance
of 7.6~Mpc) corresponding to R=8.7\,kpc (using the same de-projection as \citealt{bresolin11}). Fig~\ref{target_b}
shows the location of the target within NGC\,4258 and a zoom of the HST ACS B, V, I images. The
second source visible in these images is significantly fainter: V$=23.47\pm0.06$~mag, B-V$=-0.11\pm0.09$~mag,
V-I$=-0.15\pm0.10$~mag. The LRIS slit was oriented perpendicular to the direction between the two sources to
minimize contamination by the fainter star.

\begin{figure}
\begin{center}
  \includegraphics[scale=0.35,angle=90]{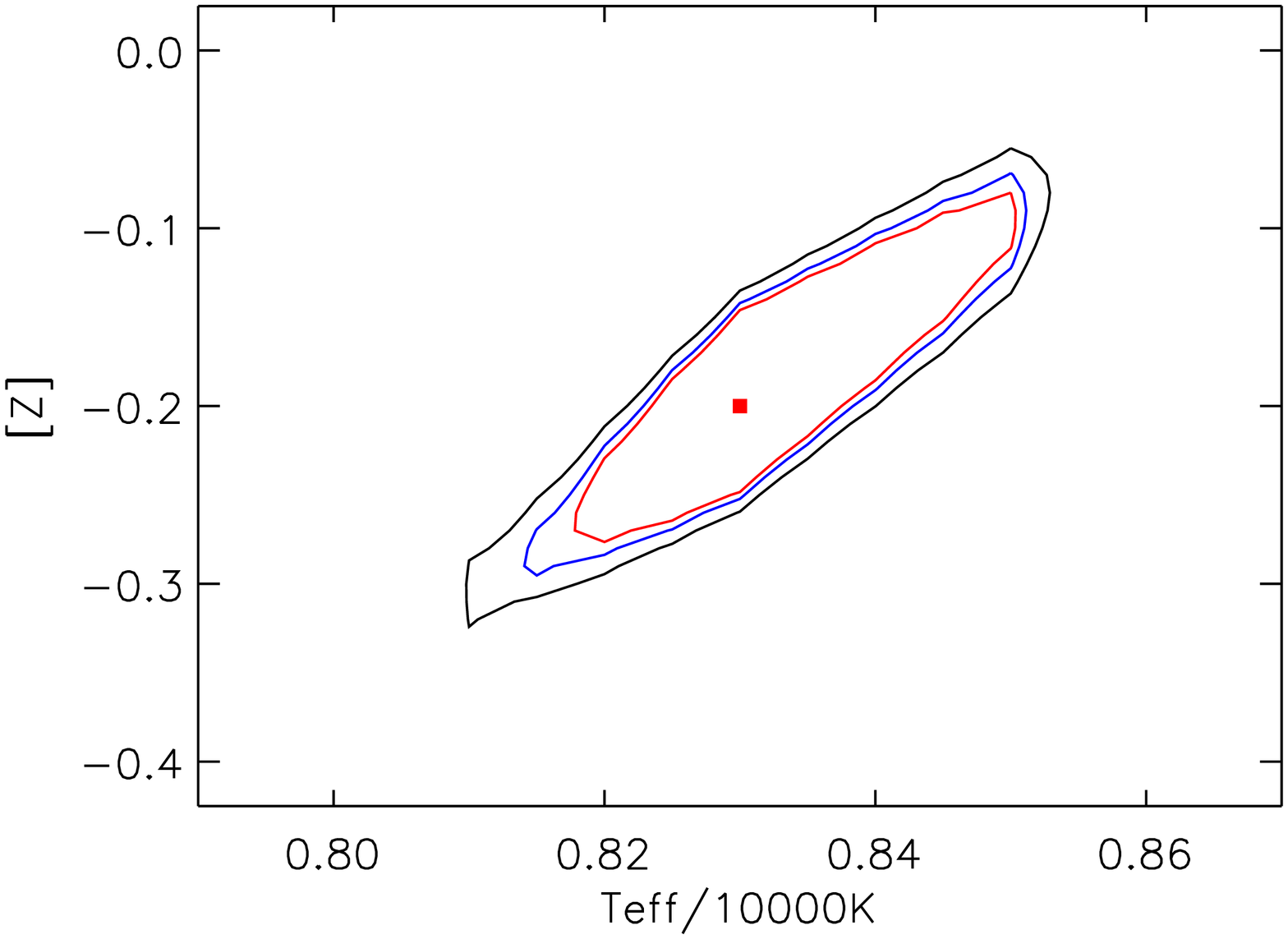}
\caption{Isocontours $\Delta \chi^{2}$  of the metallicity [Z] and effective temperature \teff fit of metal lines. 
The $\Delta \chi^{2}$ values of the isocontours are $\Delta \chi^{2}$=3 (red), 6 (blue), 12 (black), respectively.   \label{isocont}}
 \end{center}
\end{figure}

Nine MOS on-target exposures were taken each 45 minutes long. The extraction and reduction of the spectra was 
carried out in exactly the same way as described in \citet{kud12}. The final co-added and continuum normalized 
spectrum was smoothed over six pixels and has an average S/N level of 65. 

\section{Spectroscopic Analysis}

\begin{figure}
\begin{center}
  \includegraphics[scale=0.35,angle=90]{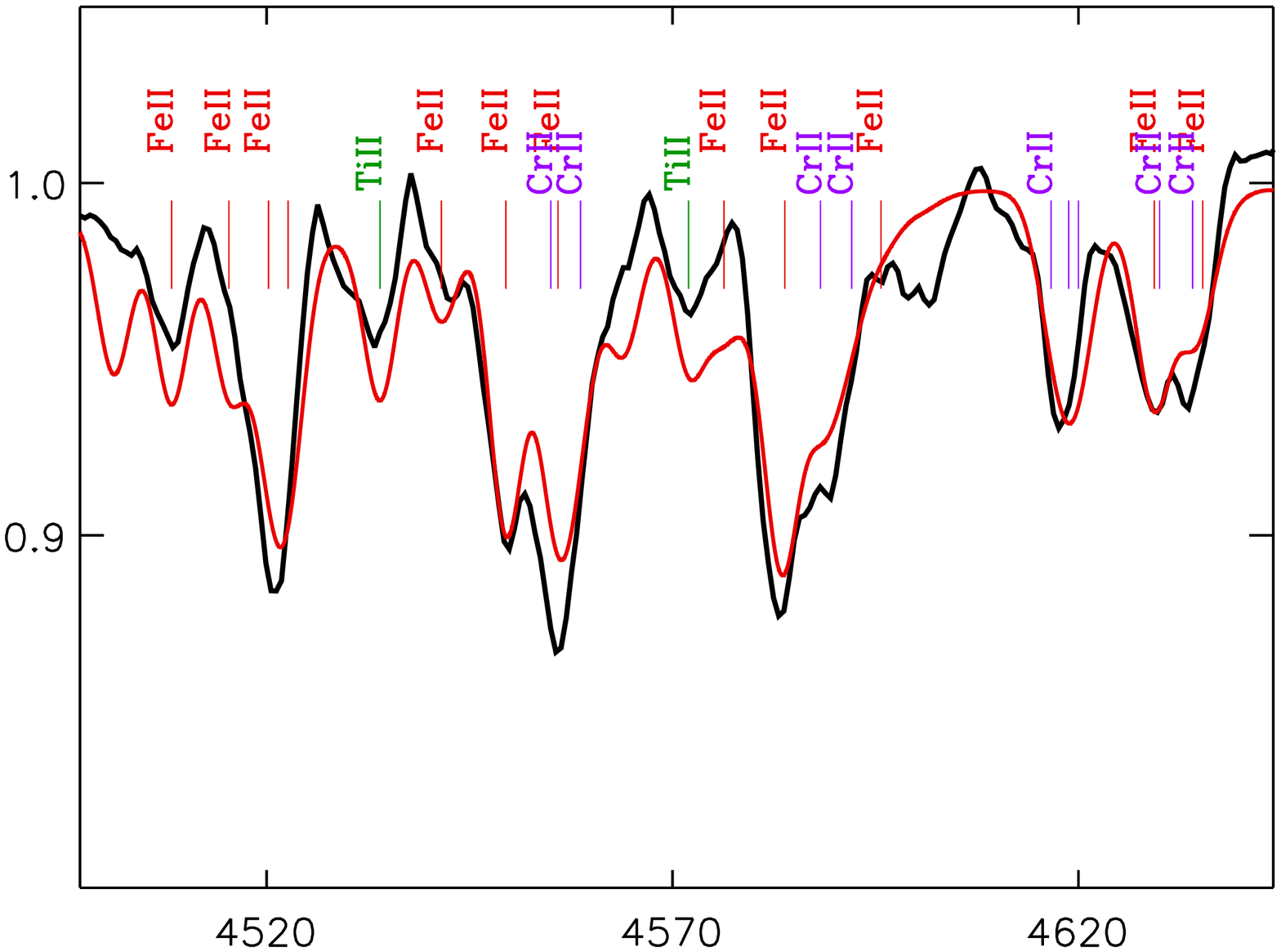}
 \includegraphics[scale=0.3,angle=90]{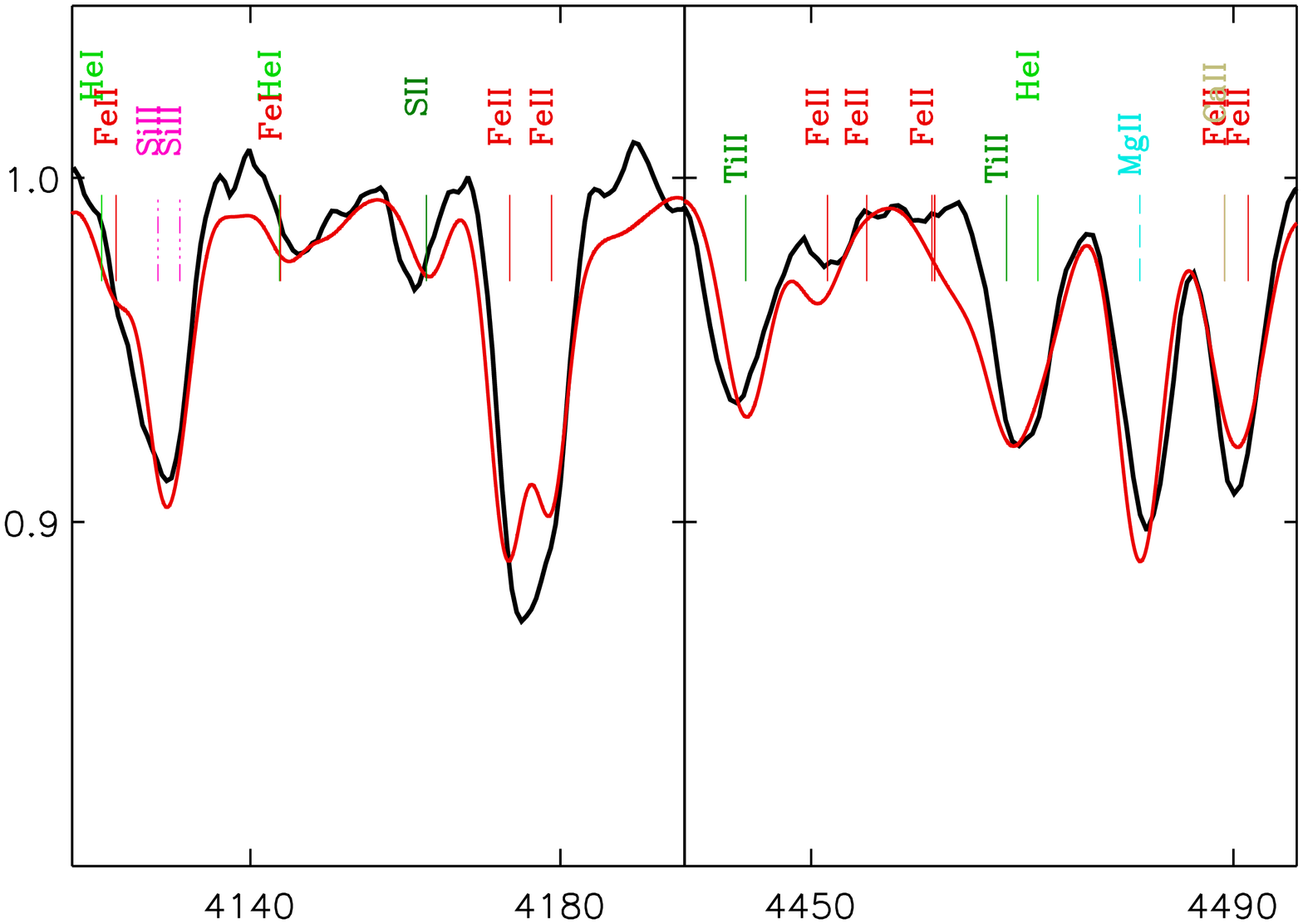} 
\caption{Observed metal line spectrum (black) in three spectral windows compared with a model spectrum (red) calculated for \teff=8300K, log g=0.85 (cgs) and [Z]=-0.15. \label{metalfit}}
 \end{center}
\end{figure}

The basis for the the quantitative determination of stellar effective temperature, gravity and metallicity is a
comprehensive grid of line-blanketed model atmospheres and very detailed NLTE line formation calculations of
normalized spectra (for details, see \citealt{kud08}, \citealt{kud12}).  Comparing calculated and observed spectra the
spectral analysis proceeds in several steps.  First, a fit curve in the (log g, T$_{\rm eff}$)-plane is constructed,
along which the models reproduce the observed Balmer lines. At every T$_{\rm eff}$ the fit of the higher Balmer lines
H$_{6,8,9,10}$ yields a value of gravity log g (see Fig~\ref{balmer}) with an accuracy better than 0.05 dex.  At lower
T$_{\rm eff}$ the log g fit-values are lower, and they are higher at higher T$_{\rm eff}$ (typical Balmer line fit
curves are shown in Kudritzki et al., 2008 or 2012). We do not use H$_{4,5}$ because these lines are contaminated by
stellar wind and \hii~region emission.  H$_{7}$ is blended by strong interstellar CaII.

In a next step, we move along this Balmer line fit curve in the (log g, T$_{\rm eff}$)-plane and compare at each
T$_{\rm eff}$ the observed and calculated spectrum of metal lines in eight spectral windows.  We split the spectrum in
spectral windows for a piecewise accurate continuum normalization and to avoid Balmer lines, nebular emission lines
and flaws in the spectrum. We then calculate a $\chi^{2}$-value as a function of logarithmic metallicity relative to
the sun [Z]=log Z/Z$_{\odot}$ and effective temperature T$_{\rm eff}$.

\begin{equation}
 \chi^{2}([Z],T_{eff}) = (S/N)^{2} \sum^{n_{pix}}_{j=1} (F_{j}^{obs} - F([Z],T_{eff})_{j}^{calc})^{2}
\end{equation}

\noindent{where S/N is the average signal-to-noise ratio per resolution element. The sum is extended over all spectral windows and
n$_{pix}$ is the sum of all pixel in all spectral windows. We then determine the minimum $\chi^{2}_{min}$ in the (log
g, T$_{\rm eff}$)-plane and calculate $\Delta \chi^{2}$ isocontours around this minimum. In order to assess the
uncertainty of this $\chi^{2}$ fitting procedure we carry out extensive Monte Carlo calculations to determine which
$\Delta \chi^{2}$ isocontour encloses 68\% of the MC solutions obtained. We find that $\Delta \chi^{2}$=3 is a
conservative estimate in reasonable agreement with statistical theory. More details describing the whole analysis
process can be found in \citet{hosek13}.}

\begin{figure}
\begin{center}
  \includegraphics[scale=0.35,angle=90]{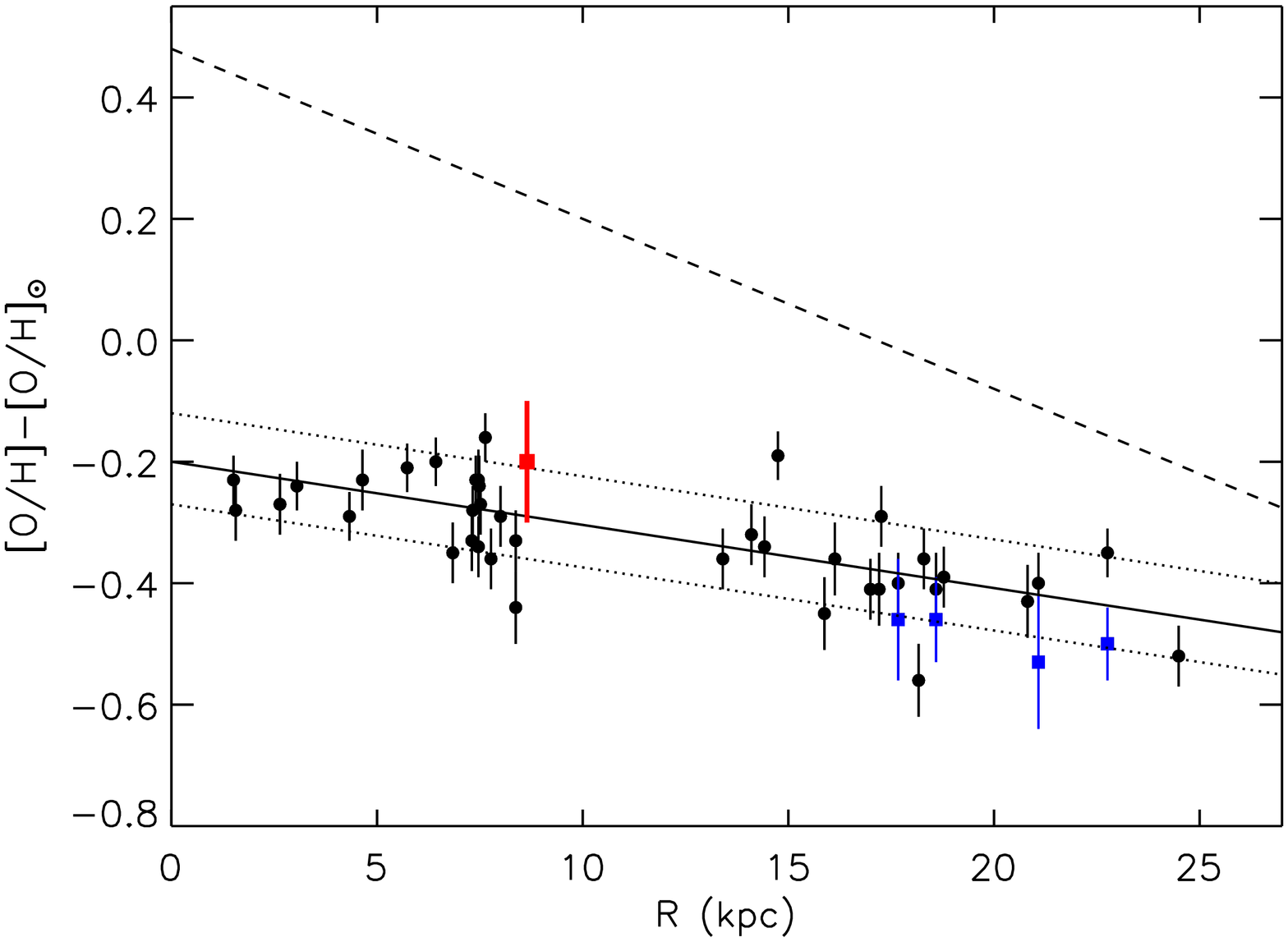}
  \includegraphics[scale=0.35,angle=90]{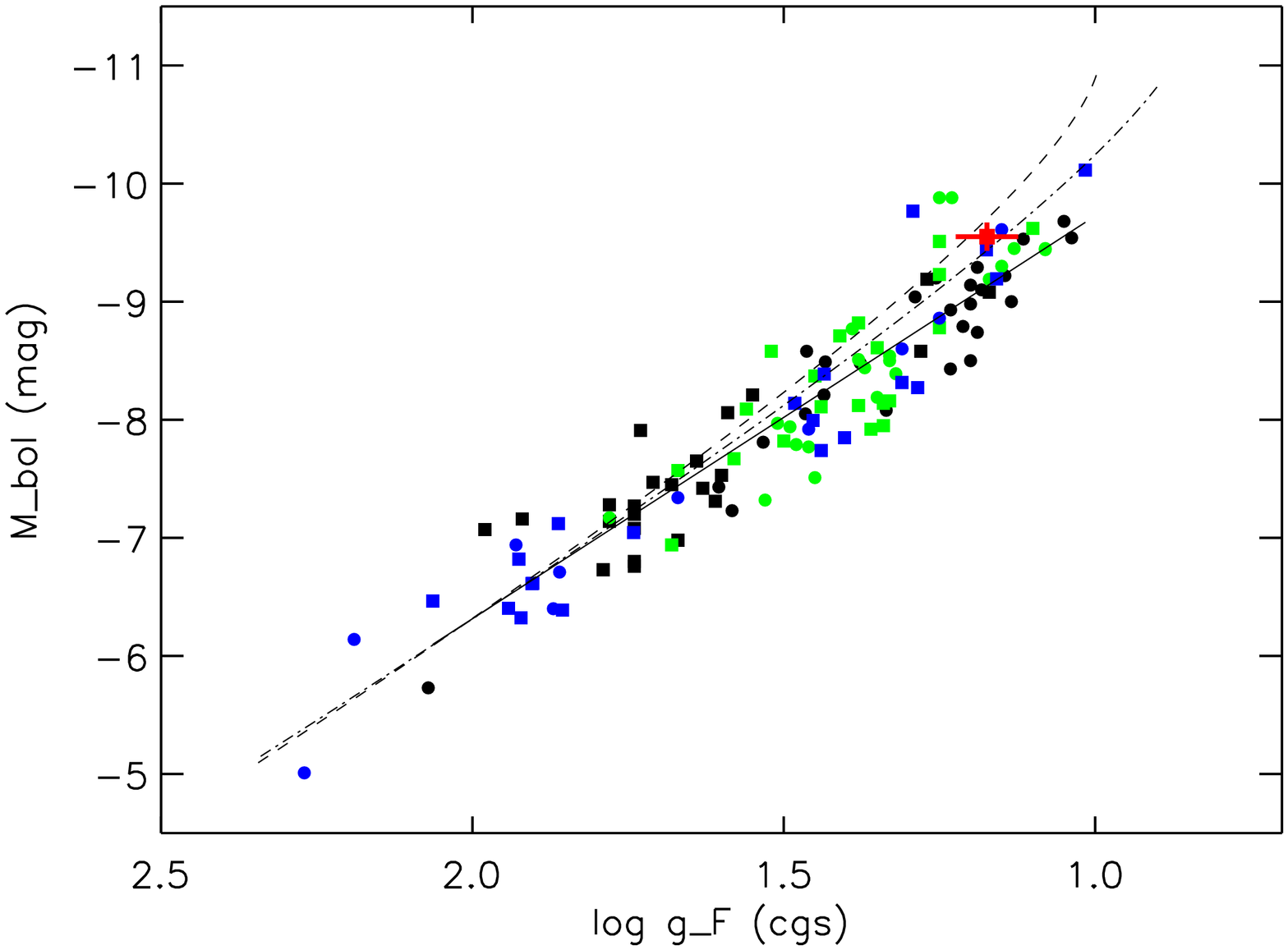}
\caption{
Top: Stellar Metallicity [Z] of the BSG (red) in NGC\,4258 shown together with the 
\hii~region oxygen abundances by \citet{bresolin11} as a function of galactocentric radius.
The oxygen abundances are normalized to the solar value (see text). Black circles and blue 
squares correspond to strong line and auroral abundances respectively. The solid line is the 
Bresolin \hii~region abundance gradient regression (with $\pm$1 $\sigma$ shown dotted), whereas the dashed line represents the 
\citet{zaritsky94} abundance gradient and metallicity used in previous work (see text). 
Bottom: Flux weighted Gravity - luminosity relationship (FGLR) of BSGs observed in 11 galaxies 
(from \citealt{kud12} and \citealt{hosek13}) together with the calibration by \citet{kud08} 
(solid) and the prediction by stellar evolution theory from evolutionary track by 
\citet{meynet05} including the effects of rotation for LMC (dashed-dotted) and SMC (dashed) 
metallicity. The BSG in NGC\,4258 is shown in red. \label{fglr}}
 \end{center}
\end{figure}

Fig~\ref{isocont} shows the isocontours in the (log g, T$_{\rm eff}$)-plane obtained for our BSG in this fitting
process. From this figure we read off an effective temperature of T$_{\rm eff}$=8300$^{+200}_{-100}$K and a
metallicity [Z]=-0.20$\pm$0.10. The gravity at the central fit point is log g = 0.85$\pm$0.05.  Fig~\ref{metalfit}
displays fits of the metal lines with a model close to the final paramters of the fit in three spectral windows.
Fig~\ref{fglr} (top) compares our stellar metallicity with the \hii~region results by \citet{bresolin11}.

Our analysis uses hydrostatic model atmospheres. This raises the question 
whether the abundance determination could be affected by a stellar wind.
The effective temperature and gravity resulting from the analysis
put the star relatively close to the Eddington-limit.
As a consequence, it may  have a strong stellar wind 
as known from other A-supergiants with similar stellar parameters 
\citep{kud99,mccarthy97}. However, detailed 
metal abundance studies using high spectral resulution and high S/N spectra
of these objects have shown that their photospheric metal lines 
and the resulting abundances are not influenced by the outer 
atmosphere effects of stellar winds (Przybilla et al., 2006, 2008).
Thus, while our low resolution spectra with the strongest Balmer 
lines H$_{\alpha,\beta}$ contaminated by \hii~emission do not allow to 
constrain the strength of the wind (note that objects of 
this type have low wind speeds of about 200 km/s 
only), we conclude that stellar wind effects are unlikely to 
affect our metallicity determination.

We can use the stellar parameters obtained in this way to calculate the intrinsic B-V and V-I colors of the BSG to
determine the reddening. We obtain E(B-V)=0.20~mag and E(V-I)=0.34~mag. With E(B-V)=0.78E(V-I), we get E(B-V)=0.26~mag
for the E(V-I) value determined. Thus, we adopt E(B-V)=0.23$\pm$0.03~mag for the reddening and with R$_{V}$=3.2
calculate A$_{V}$=0.73$\pm$0.10~mag for the extinction. With the distance modulus of m-M$=29.40\pm0.07$~mag to
NGC\,4258 and a bolometric correction from the model calculations of BC=-0.006~mag we finally obtain
M$_{V}$=M$_{bol}$=-9.55$\pm$0.12~mag for the visual and bolometric magnitudes, respectively. The quoted
  uncertainty includes contributions from the photometry, extinction correction and distance modulus.

With stellar temperature and gravity determined BSGs can also be used as distance indicators using the 
flux-weighted gravity--luminosity relationship (FGLR) introduced by \citet{kud03} and \citet{kud08}.
The FGLR relates the flux-weighted gravity ($g_F\,\equiv\,g/{T^4}_{\rm eff}, T_{\rm eff}$ in units of 10$^{4}$K) 
of BSGs to their absolute bolometric magnitude M$_{bol}$

\begin{equation}
 M_{\rm bol}\,=\,a (\log\,g_F\,-\,1.5)\,+\,b
\end{equation}

\noindent{with $a$ = 3.41 and $b$ = -8.02 as determined by \citet{kud08}.
BSGs form such a relationship, because they evolve accross the HRD at roughly constant luminosity and mass.
As a consequemce, $g_F$ remains constant during the horizontal HRD evolution independent of temperature, 
while at the same time the luminosity is a strong function of stellar mass and, therefore, $g_F$ .
Distance determinations using the FGLR have already been carried for a number of galaxies 
(see \citealt{kud12} and \citealt{hosek13} and references, therein).}

With only one object at this point no independent determination of the distance to NGC\,4258 is possible. 
However, we can at least discuss the one observed BSG in the maser galaxy in relation to the objects already studied in other 
galaxies. This is done in Fig~\ref{fglr}. The result is encouraging. While the object lies somewhat above 
the calibration relationship it is still within the observed scatter. Moreover, at the high luminosity 
end there is the prediction by stellar evolution theory of a possible curvature of the FGLR, for which 
we may already see an indication in the data observed (for a more detailed discussion, see \citealt{hosek13}). 
Of course, this needs to be investigated by adding more BSGs belonging to NGC\,4258 to the plot and by fitting a 
distance with less luminous objects. This work is presently under way.

\section{Discussion} 

In our spectroscopic analysis of a BSG at a galactocentric distance of 8.7\,kpc we have found a stellar metallicity of
-0.2 dex relative to the solar value. Given the uncertainty of 0.1 dex, this is in reasonable
agreement with the recent \hii~region work by \citet{bresolin11} which obtained a value of $-0.29\pm0.08$\,dex relative to
solar at this galactocentric distance. While this is only the study of one object, it is already a strong confirmation
that the stellar metallicity of the young population in the disk of NGC\,4258 is very similar to that of the LMC, for
which \citet{romaniello08} have found an average value of [Z]$_{\rm LMC}$=-0.33 from a spectroscopic analysis of
Cepheid variables (in perfect agreement with the \hii~region value of \citealt{bresolin11}). Clearly, only by a
study of a larger sample of BSGs in NGC\,4258 covering a range of galactocentric distances will we be able to
confirm whether the metallicity gradient is as shallow as found by \citet{bresolin11} and the average metallicity
clearly lower than solar. This work is presently under way.

The quantitative spectroscopy of the one object investigated so far has resulted in a bolometric magnitude 
and flux-weighted gravity in agreement with the FGLR of BSGs in galaxies. This is an encouraging first step 
towards a new calibration of this relationship in the maser galaxy NGC\,4258 as a new anchor point of the 
extragalctic distance scale.


\acknowledgments This work was supported by the National Science Foundation under grant AST-1008798 to RPK and FB.
LMM acknowledges support by NASA through grants HST-GO-10802.14 and HST-GO-11570.09 from the Space
Telescope Science Institute, which is operated by the Association of Universities for Research in Astronomy, Inc.,
under NASA contract NAS 5-26555; and by the Department of Physics \& Astronomy at Texas A\&M University through
faculty startup funds and the Mitchell-Heep-Munnerlyn Endowed Career Enhancement Professorship in Physics or
Astronomy.

The data presented in this work were obtained at the W.M. Keck Observatory, which is operated as a scientific 
partnership among the California Institute of Technology, the University of California and the National Aeronautics and 
Space Administration. The Observatory was made possible by the generous financial support of the W.M. Keck Foundation.

Based on observations made with the NASA/ESA Hubble Space Telescope, obtained from the Data Archive at the Space
Telescope Science Institute, which is operated by the Association of Universities for Research in Astronomy, Inc.,
under NASA contract NAS 5-26555. These observations are associated with programs \#10802 and \#11570.

The authors wish to recognize and acknowledge the very significant cultural role and reverence that the summit of 
Mauna Kea has always had within the indigenous Hawaiian community.  We are most fortunate to have the opportunity to 
conduct observations from this mountain.




{\it Facilities:} \facility{Keck (LRIS)}, \facility{HST (ACS)}.

\clearpage




\clearpage









\begin{thebibliography}{}

\bibitem[Allende Prieto et al.(2001)]{allende01} Allende Prieto, C., Lambert, D.~L., \& Asplund, M.\ 2001, \apjl, 556, L63 
\bibitem[Asplund et al.(2009)]{asplund09} Asplund, M., Grevesse, N., Sauval, A.~J., \& Scott, P.\ 2009, ARA\&A, 47, 481 
\bibitem[B{\`e}land et al.(1988)]{beland88} B{\`e}land, S., Boulade, O., \& Davidge, T.\ 1988, Bulletin d'information du telescope Canada-France-Hawaii, 19, 16 
\bibitem[Bresolin et al.(2006)]{bresolin06} Bresolin, F., Pietrzy{\'n}ski, G., Urbaneja, M.~A., et al.\ 2006, \apj, 648, 1007 
\bibitem[Bresolin et al.(2007)]{bresolin07} Bresolin, F., Urbaneja, M.~A., Gieren, W., Pietrzy{\'n}ski, G., \& Kudritzki, R.-P.\ 2007, \apj, 671, 2028 
\bibitem[Bresolin et al.(2009)]{bresolin09} Bresolin, F., Gieren, W., Kudritzki, R.-P., et al.\ 2009, \apj, 700, 309 
\bibitem[Bresolin(2011)]{bresolin11} Bresolin, F.\ 2011, \apj, 729, 56 
\bibitem[Cardelli et al.(1989)]{cardelli89} Cardelli, J.~A., Clayton, G.~C., \& Mathis, J.~S.\ 1989, \apj, 345, 245 
\bibitem[Evans et al.(2007)]{evans07} Evans, C.~J., Bresolin, F., Urbaneja, M.~A., et al.\ 2007, \apj, 659, 1198 
\bibitem[Freedman et al.(2001)]{freedman01} Freedman, W.~L., Madore, B.~F., Gibson, B.~K., et al.\ 2001, \apj, 553, 47 
\bibitem[Gerke et al.(2011)]{gerke11} Gerke, J.~R., Kochanek, C.~S., Prieto, J.~L., Stanek, K.~Z., \& Macri, L.~M.\ 2011, arXiv:1103.0549 
\bibitem[Herrnstein et al.(1999)]{herrnstein99} Herrnstein, J.~R. et al.\ 1999, Nature, 400, 539 
\bibitem[Hosek et al.(2013)]{hosek13} Hosek Jr., M.~W., Kudritzki, R.~P., Bresolin et al., \ 2013, \apj, submitted 
\bibitem[Humphreys et al.(2008)]{humphreys08} Humphreys, E.~M.~L., Reid, M.~J., Greenhill, L.~J., Moran, J.~M., 
\& Argon, A.~L.\ 2008, \apj, 672, 800 
\bibitem[Humphreys et al.(2013)]{humphreys13} Humphreys, E.~M.~L., Reid, M.~J., Greenhill, L.~J., Moran, J.~M., 
\& Argon, A.~L.\ 2013, \apj, 775, 13 
\bibitem[Kennicutt et al.(1998)]{kennicutt98} Kennicutt, R.~C., Jr., Stetson, P.~B., Saha, A., et al.\ 1998, \apj, 498, 181 
\bibitem[Kewley \& Ellison(2008)]{kewley08} Kewley, L.~J., \& Ellison, S.~L.\ 2008, \apj, 681, 1183 
\bibitem[Kudritzki et al.(2012)]{kud12} Kudritzki, R.~P., Urbaneja, M.~A., Gazak, Z., et al.\ 2012, \apj, 747, 15 
\bibitem[Kudritzki et al.(2008)]{kud08} Kudritzki, R.~P., Urbaneja, M.~A., Bresolin, F., et al.\ 2008, \apj, 681, 269 
\bibitem[Kudritzki et al.(2003)]{kud03} Kudritzki, R.~P., Bresolin, F., \& Przybilla, N.\ 2003, \apjl, 582, L83 
\bibitem[Kudritzki et al.(1999)]{kud99} Kudritzki, R.~P., Puls, J., Lennon, D.~J., et al.\ 1999, \aap, 350, 970
\bibitem[Leavitt \& Pickering (1912)]{leavitt12} Leavitt, H.~S \& Pickering, E.~C.\ 1912, Harvard College Observatory Circular 173, 1
\bibitem[Macri et al.(2006)]{macri06} Macri, L.~M., Stanek, K.~Z., Bersier, D., Greenhill, L.~J., \& Reid, M.~J.\ 2006, \apj, 652, 1133 
\bibitem[Majaess et al. (2011)]{majaess11} Majaess, D., Turner, D., \& Gieren, W. \ 2011, \apj, 741, L36
\bibitem[McCarthy et al. (2006)]{mccarthy97} McCarthy, J.~K., Kudritzki, R.~P., Lennon, D.,~J., et al.\ 1997, \apj, 482, 757
\bibitem[McCommas et al.(2009)]{mccommas09} McCommas, L.~P., Yoachim, P., Williams, B.~F., et al.\ 2009, \aj, 137, 4707 
\bibitem[Meynet \& Maeder\/(2005)]{meynet05} Meynet, G. \& Maeder, A. \ 2005, \aap, 429, 581
\bibitem[Oke(1990)]{oke90} Oke, J.~B.\ 1990, \aj, 99, 1621 
\bibitem[Oke et al.(1995)]{oke95} Oke, J.~B., Cohen, J.~G., Carr, M., et al.\ 1995, \pasp, 107, 375 
\bibitem[Pietrzy{\'n}ski et al. (2013)]{pietrzynski13} Pietrzy{\'n}ski, G. et al. \ 2013, Nature, 495, 76
\bibitem[Przybilla et al.(2006)]{przybilla06} Przybilla, N., Butler, K., Becker, S.~R., \& Kudritzki, R.~P.\ 2006, \aap, 445, 1099 
\bibitem[Przybilla et al.(2008)]{przybilla08} Przybilla, N., Butler, K., \& Kudritzki, R.-P.\ 2008, The Metal-Rich Universe, ed. 
G. Israelian, \& G. Meynet (Cambridge: Cambridge University Press), 332
\bibitem[Riess et al.(2009)]{riess09} Riess, A.~G., Macri, L., Casertano, S., et al.\ 2009, \apj, 699, 539
\bibitem[Riess et al.(2011)]{riess11} Riess, A.~G., Macri, L., Casertano, S., et al.\ 2011, \apj, 730, 119 
\bibitem[Romaniello et al.(2008)]{romaniello08} Romaniello, M., Primas, F., Mottini, M., et al.\ 2008, \aap, 488, 731 
\bibitem[Schlegel et al.(1998)]{schlegel98} Schlegel, D.~J., Finkbeiner, D.~P., \& Davis, M.\ 1998, \apj, 500, 525 
\bibitem[Shappee \& Stanek (2011)]{shappee11} Shappee, B.~J, \& Stanek, K.~Z., 2011, \apj, 733, 124
\bibitem[Storm et al.(2011)]{storm11} Storm, J., Gieren, W., Fouque, P., et al.\ 2011, \aap, 534, 95
\bibitem[U et al.(2009)]{u09} U, V., Urbaneja, M.~A., Kudritzki, R.-P., et al.\ 2009, \apj, 704, 1120 
\bibitem[Urbaneja et al.(2008)]{urbaneja08} Urbaneja, M.~A., Kudritzki, R.-P., Bresolin, F., et al.\ 2008, \apj, 684, 118 
\bibitem[Zaritsky et al.(1994)]{zaritsky94} Zaritsky, D., Kennicutt, R.~C., Jr., \& Huchra, J.~P.\ 1994, \apj, 420, 87 

\end{thebibliography}
\end{document}